# Optical signatures of low spin $Fe^{3+}$: a new probe for the spin state of bridgmanite and post-perovskite

**Sergey S. Lobanov[1,2]\*, Han Hsu[3]†, Jung-Fu Lin[4,5], Takashi Yoshino[6], Alexander F. Goncharov[1,7]**

[1]Geophysical Laboratory, Carnegie Institution of Washington, Washington, DC 20015, USA

[2]Sobolev Institute of Geology and Mineralogy Siberian Branch Russian Academy of Sciences, 3 Pr. Ac. Koptyga, Novosibirsk 630090, Russia

[3]Department of Physics, National Central University, Taoyuan City 32001, Taiwan

[4]Department of Geological Sciences, Jackson School of Geosciences, The University of Texas at Austin, Austin, TX 78712, USA

[5]Center for High Pressure Science and Technology Advanced Research (HPSTAR), Shanghai 201900, China

[6]Institute for Planetary Materials, Okayama University, Misasa, Tottori 682-0193, Japan

[7]Key Laboratory of Materials Physics, Institute of Solid State Physics CAS, Hefei 230031, China

\*E-mail: slobanov@carnegiescience.edu; slobanov@igm.nsc.ru

†E-mail: hanhsu@ncu.edu.tw

**Abstract**

Iron spin transition directly affects properties of lower mantle minerals and can thus alter geophysical and geochemical characteristics of the deep Earth. While the spin transition in ferropericlase has been vigorously established at P ~ 60 GPa and 300 K, experimental evidence for spin transitions in other rock-forming minerals, such as bridgmanite and post-perovskite, remains controversial. Multiple valence, spin, and coordination states of iron in bridgmanite and post-perovskite are difficult to resolve with conventional spin-probing techniques. Optical spectroscopy, on the other hand, is sensitive to high/low spin ferrous/ferric iron at different sites; thus, it can be a powerful probe for spin transitions. Here we establish the optical signature of low spin $Fe^{3+}O_6$, a plausible low spin unit in bridgmanite and post-perovskite, by optical absorption experiments in diamond anvil cells. We show that the optical absorption of $Fe^{3+}O_6$ in NAL (new aluminous phase) is very sensitive to the iron spin state and represents a model behavior of bridgmanite and post-perovskite in the deep lower mantle across a spin transition.

Specifically, an absorption band centered at ~ 19000 cm$^{-1}$ is characteristic of the $^2T_{2g} \rightarrow {}^2T_{1g}$ ($^2A_{2g}$) transition in low spin $Fe^{3+}$ in NAL at 40 GPa. This new spectroscopic information constrains the crystal field splitting energy of low spin $Fe^{3+}$ to ~ 22200 cm$^{-1}$ which we also independently confirm by our first-principles calculations. Together with available information on the electronic structure of $Fe^{3+}O_6$-compounds, we constrain the spin-pairing energy of $Fe^{3+}$ in an octahedral field to ~ 20000-23000 cm$^{-1}$. This implies that octahedrally-coordinated $Fe^{3+}$ in bridgmanite is low spin at P > ~ 40 GPa.

**Introduction**

It has been reported that the spin transition of iron in lower mantle minerals can greatly impact physical and chemical properties of the deep Earth[1-3]. For example, high-pressure experiments at room temperature have identified anomalies in elastic and transport properties of ferropericlase (Fp) across the spin transition at P ~ 60 GPa (Refs.[4-7]) which may help elucidate seismic signatures and geodynamic processes of the lower mantle[6-9]. It has also been proposed that the spin transition can affect the iron partitioning coefficient between mantle phases due to the associated decrease in iron atomic volume[1, 10]. Yet, there are many unknowns in the exact role of spin transitions in the deep Earth.

Pressure-induced spin transitions have now been identified in multiple minerals that may be present in the lower mantle[2, 3]. However, the spin state of bridgmanite (Bdgm), the most abundant lower mantle phase, remains controversial in the mechanism, pressure, and compositional dependence[11-20]. The main reason for this controversy arises from its crystal chemical complexity, in which the iron can be distributed between two non-equivalent crystallographic positions: large pseudo-dodecahedral (*A*) sites and relatively small octahedral (*B*) sites[21]. Ferric iron ($Fe^{3+}$) enters both *A* and *B* sites via charge-coupled substitution mechanism ($Fe^{3+} + Al^{3+} \rightarrow Si^{4+} + Mg^{2+}$ or $2\ Fe^{3+} \rightarrow Si^{4+} + Mg^{2+}$), while ferrous iron ($Fe^{2+}$) predominantly substitutes for $Mg^{2+}$ at the *A* site[22]. Importantly, aluminum admixture to the *B* site reduces its capacity to host ferric iron[18, 23].

Theoretical investigations indicate that the *B*-site $Fe^{3+}$ undergoes a transition from the high spin (HS) to the low spin (LS) state, while *A*-site iron remains in the HS state, regardless of its valence[20, 24-28]. This theory-based view has found experimental support[15-17], but is not yet widely accepted[14, 18]. Earlier studies of the Bdgm spin state via x-ray emission spectroscopy (XES)[11, 12], a technique that is not site- and valence-specific, yielded an averaged spin moment for all iron ions at all possible crystallographic sites. Mössbauer spectroscopy (traditional or synchrotron) can be used to infer the iron oxidation/spin states as well as the site occupancies, but the relation of hyperfine parameters to various iron states is not straightforward and, as a result, the interpretation of Mössbauer spectra is generally inconclusive[29]. Combined XES and Mössbauer studies helped to partially resolve the problem[15, 16, 19, 30], but are yet inefficient in probing a broad range of chemical compositions because it requires extraordinary access to synchrotron x-ray sources.

Alternatively, the spin and valence states of iron in a host mineral may be probed by optical absorption methods[5, 10, 31, 32] which are fast (seconds to minutes vs hours for XES and Mösbauer) and do not require synchrotron sources. Unfortunately, previous optical studies of Bdgm[33, 34] did not recognize the complexity of iron states in the mineral and lacked understanding of how the optical spectra of Bdgm would change across various spin transition scenarios. Consequently, optical studies of the Bdgm spin state were discontinued.

The goal of this work is to address optical absorption and crystal field characteristics of mixed-valence iron-bearing minerals, such as Bdgm, across the spin transition. To this end, we use iron-bearing new aluminous phase (NAL) as a representative model of such compounds with a spin transition in $Fe^{3+}O_6$ at ~ 40 GPa (Ref.[35]). Recently, a Mössbauer study of iron-bearing NAL has shown that ferric and ferrous iron are distributed between two edge-sharing crystallographic sites $B$ and $C$ at ambient conditions[35]. The trigonal prismatic site $B$ hosts both $Fe^{2+}$ and $Fe^{3+}$, while the octahedral site $C$ accommodates $Fe^{3+}$ only. A spin transition in $Fe^{3+}$ at the $C$ site occurs at ~ 40 GPa and 300 K, while the $B$ sites remains in HS up to at least 80 GPa (Ref.[35]). Remarkably, the proposed spin transition scenario for the octahedrally-coordinated $Fe^{3+}$ in NAL is analogous to that in Bdgm[15, 16, 25-27], hence, changes in the optical absorption of NAL and Bdgm across their respective spin transition must be similar, in accordance with the Tanabe-Sugano diagram for $d^5$ elements. Other than the aforementioned crystal-chemical similarities with Bdgm, NAL is an abundant mineral in subducting slabs with an estimated content of up to 20 % at $P$ = 15-50 GPa (Refs.[36]) and is a characteristic mineral inclusion in lower mantle diamonds[37]. Because the spin transition likely affects the partitioning of iron in the NAL-bearing lower mantle, the iron content of NAL inclusions in natural diamonds may be used to infer the $P$-$T$ parameters of the entrapment, similarly to the majorite-garnet composition that has been used to constrain the origin depth of sublithospheric diamonds. The first step towards such a thermobarometer is to characterize the crystal field of NAL across the spin transition.

Here we report room temperature visible and near-infrared (IR) absorption of iron-bearing NAL phase at 0-57 GPa. We confirm the spin transition in $Fe^{3+}$ at the *C* site at 30-47 GPa (Ref.[35]) and, for the first time, characterize the crystal field of LS ferric iron, establishing the base for optical studies of the spin transition in iron-bearing mixed-valence minerals at lower-mantle pressures. The measured crystal field splitting energy is in excellent agreement with our first-principles calculations based on density functional theory (DFT).

**Methods**

Single crystals of $Na_{0.71}Mg_{2.05}Al_{4.62}Si_{1.16}Fe^{2+}_{0.09}Fe^{3+}_{0.17}O_{12}$ NAL phase (from the same sample capsule of NAL used in a previous study[35]) were loaded in symmetrical DACs with 300 μm diamond culets. Re gaskets were indented (~ 45 μm thick) and laser-drilled in the center of the indentation to create a sample chamber (~ 90-120 μm in diameter). NAL crystals (40-70 μm wide and 10-20 μm thick) were positioned at the center of the cavity with several ruby chips put 10-30 μm aside from the sample for pressure determination. Argon pressure medium was gas-loaded at 0.2 GPa.

Absorption spectra were collected using a custom microscope with all-reflective relay optics in the UV-visible (10000-25000 $cm^{-1}$) and near-mid IR (3000-11000 $cm^{-1}$) range. For the UV-visible range we used a fiber-coupled halogen-deuterium light source focused to a ~ 50 μm spot on the sample. The central portion of the transmitted light (~ 20 μm) was selected by a confocal aperture and passed to the spectrograph (Acton Research Corporation SpectraPro 500-i) equipped with a 300 grooves/mm grating and a CCD detector operated at 235 K. For the IR range we used a Varian Resolution Pro 670-IR Fourier-transform spectrometer with quartz and KBr beamsplitters. The optical setup was described in detail by Goncharov et al. [38]. After the collections in UV-visible and IR ranges the spectra were stitched together to evaluate the sample optical absorbance: $A(v) = -log_{10}(I_{sample} - I_{bckg})/(I_{reference} - I_{bckg})$, where $I_{sample}$ is the intensity of light transmitted through the sample, $I_{reference}$ is the intensity of light passed through the pressure medium, and $I_{bckg}$ is the background reading. All spectra were collected at room temperature. Absorption bands were fitted with a Gaussian function after a linear baseline subtraction.

First-principles calculations based on DFT were performed with the Quantum ESPRESSO codes[39]. We used 63-atom supercells (4×4×4 **k**-point mesh) of $NaMg_2(Al_5,Si)O_{12}$ and $NaMg_2(Al_{4.67},Fe_{0.33},Si)O_{12}$ to simulate iron-free and iron-bearing ($Fe^{3+}$ in the $C$ site) samples, respectively. The fully optimized atomic structures will be reported elsewhere. The PBE-type generalized gradient approximation (PBE-GGA)[40] and ultrasoft pseudopotentials (USPPs) generated with the Vanderbilt method[41] were adopted in our calculations. The USPPs of Mg, Al, Si, O, and Fe are detailed in Ref.[24]; the USPP of Na is reported in Ref.[42].

**Experimental Results**

Tanabe-Sugano diagrams predict a $^5T_{2g} \to {}^5E_g$ spin-allowed crystal field absorption band for HS $Fe^{2+}$ and zero spin-allowed transitions for HS $Fe^{3+}$ in NAL phase. Spin-forbidden bands may also be present in the spectra as these may be intensified by magnetic interactions of $Fe^{2+}$ and $Fe^{3+}$ in adjacent edge-sharing $B$ and $C$ sites[10]. Additionally, a $Fe^{2+}$-$Fe^{3+}$ intervalence charge transfer (CT) may be expected due to the edge-sharing character of the $B$ and $C$ sites which enhances exchange interactions between the sites[10]. LS $Fe^{3+}$ ($^2T_{2g}$ ground state) has 4 spin-allowed crystal field transitions; thus, the spin crossover in Fe-bearing NAL phase should be optically apparent, consistent with our visual observations of NAL single crystal compressed to ~ 50 GPa (Fig. 1).

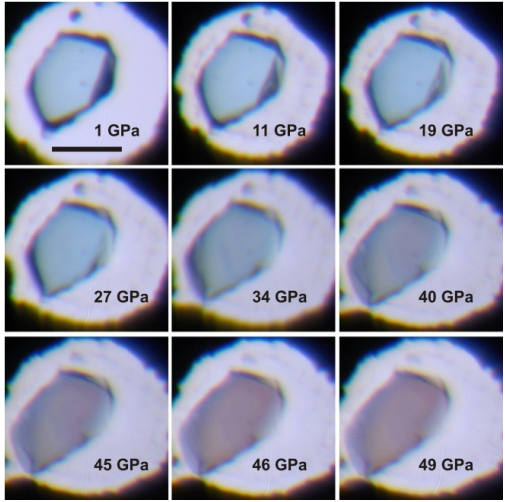

**Figure 1.** High-pressure optical images of $Na_{0.71}Mg_{2.05}Al_{4.62}Si_{1.16}Fe^{2+}_{0.09}Fe^{3+}_{0.17}O_{12}$ NAL single crystal compressed to ~ 50 GPa at room temperature taken at identical illumination and image processing conditions. A change in color is observed at ~ 40 GPa indicating a pressure-induced spin transition. The black bar corresponds to 50 μm.

At near-ambient conditions, the blue color of iron-bearing NAL phase is due to a single, broad absorption feature at ~ 14500 cm$^{-1}$ (Fig. 2A). Typical energies of the HS $Fe^{2+}$ crystal field band ($^5T_{2g} \rightarrow {}^5E_g$) in a 6-fold coordination (trigonal prismatic site *B*) are 10000-12000 cm$^{-1}$, while $Fe^{2+}$-$Fe^{3+}$ CT occurs at 12000-16000 cm$^{-1}$ (Refs.[10, 43]). For this reason we assume that the maximum of absorption intensity at 14500 cm$^{-1}$ arises from the CT band, obscuring the $^5T_{2g} \rightarrow {}^5E_g$ band which we presume also contributes to the spectrum at 10000-15000 cm$^{-1}$. The broadness of the main absorption feature also relates the band at ~ 14500 cm$^{-1}$ to the $Fe^{2+}$-$Fe^{3+}$ CT as large widths (3000-6500 cm$^{-1}$) have been diagnostic of the CT bands[43].

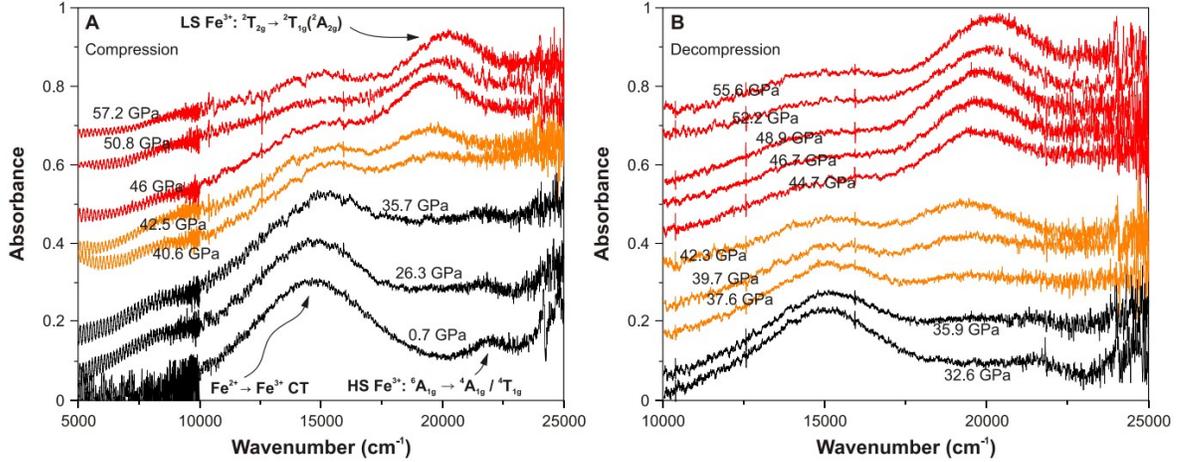

**Figure 2.** Absorption spectra of $Na_{0.71}Mg_{2.05}Al_{4.62}Si_{1.16}Fe^{2+}_{0.09}Fe^{3+}_{0.17}O_{12}$ NAL upon compression (**A**) and decompression (**B**). Spectra are offset vertically by ~ 0.1-0.2 for clarity. Significant changes are observed at P ~ 40 GPa: (i) a new band appears at ~ 19200 cm$^{-1}$ and blue-shifts with increasing pressure; (ii) the $Fe^{2+}$-$Fe^{3+}$ charge transfer (CT) softens and decreases in intensity. All transformations are reversible on decompression (**B**), as is expected for a pressure-induced spin transition.

The CT band blue-shifts in the 0-40 GPa pressure range with $dv/dP$ = 23 cm$^{-1}$/GPa (Fig. 3). Upon compression to 40-46 GPa, the CT band softens by ~ 400 cm$^{-1}$ and maintains an approximately constant frequency with further pressure increase. At P < 40 GPa, the gradual decrease in intensity is associated with the sample thinning due to compression, while at P ~ 40 GPa the CT band intensity reduces abruptly (Fig. 2A). Likewise, an increase in the CT band intensity is observed on decompression upon the LS-HS transition while the change in sample thickness is small (Fig. 2B). We propose that the magnetic coupling in the $Fe^{2+}$-$Fe^{3+}$ pair is suppressed across the HS to LS transition as the number of unpaired electrons in $Fe^{3+}$ is reduced from five (HS) to one (LS). Similar intensity variations of the CT band may take place in Bdgm across its spin transition as well as in other iron-bearing mixed-valence compounds, such as post-perovskite (Ppv), increasing the minerals' transparency and, hence, their radiative thermal conductivity. The variations in CT band intensity and spectral position are consistent with the change in color that is seen at ~ 40 GPa (Fig. 1) and, together, represent spectroscopic evidence for the pressure-induced HS-LS transition in NAL.

Another interesting absorption feature of HS NAL is a weak band at 22000 cm$^{-1}$ that shows no apparent pressure-induced shift (Figs. 2 and 3). We propose that this band is due to the $^6A_{1g} \rightarrow {}^4A_{1g}$ or $^6A_{1g} \rightarrow {}^4T_{1g}$ spin-forbidden transitions in HS Fe$^{3+}$ as these do not depend on the crystal field intensity (Fig. 4). The definitive choice between these two assignments cannot be made at this time as the Racah parameters derived (B = 675 or 555 cm$^{-1}$ for the $^4A_{1g}$ and $^4T_{1g}$ excited states, respectively) are both consistent with the reported values of B = 540-650 cm$^{-1}$ for HS Fe$^{3+}$ (Refs.[10, 44]). We note, however, that an absorption band at ~ 22000 cm$^{-1}$ has been diagnostic of HS Fe$^{3+}$ ions in minerals and is usually assigned to the $^6A_{1g} \rightarrow {}^4A_{1g}$ transition such as in Fe$^{3+}$-doped Al$_2$O$_3$ (Ref.[10]).

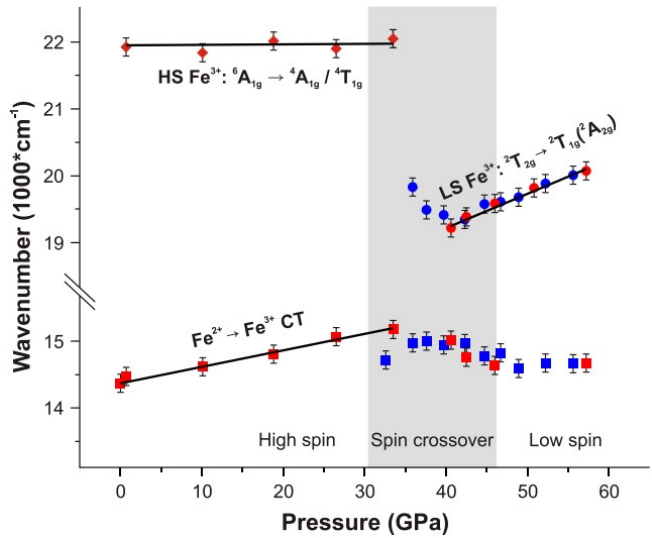

**Figure 3.** Spectral positions of the absorption bands in NAL phase (Na$_{0.71}$Mg$_{2.05}$Al$_{4.62}$Si$_{1.16}$Fe$^{2+}_{0.09}$Fe$^{3+}_{0.17}$O$_{12}$) at high pressure. Red and blue symbols show compression and decompression data, respectively. Solid lines are linear fits to the positions of the bands. The grey region corresponds to the spin transition pressure of a similar NAL sample studied by synchrotron nuclear forward scattering and x-ray diffraction[35]. The error bars are 200 cm$^{-1}$ and are due to uncertainties in spectra deconvolution.

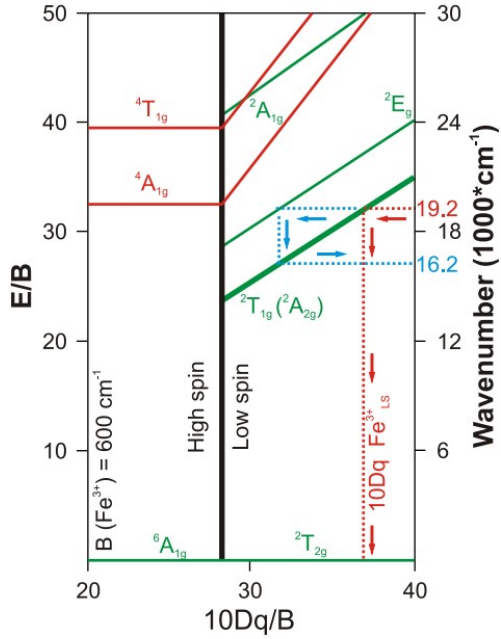

**Figure 4.** Tanabe-Sugano diagram for *d5* elements in an octahedral crystal filed. Green solid lines and labels show spin-allowed transitions and corresponding electronic states. Red solid lines are $^4A_{1g}$ and $^4T_{1g}$ spin-forbidden excited states. Blue dotted line shows that if the band at 19220 cm$^{-1}$ is due to the $^2E_g$ spin-allowed excited state, then the $^2T_{2g} \rightarrow {}^2T_{1g}({}^2A_{2g})$ band would be expected at ~ 16200 cm$^{-1}$, which is not supported by the absorption data (Fig. 3). Therefore, we assign the 19220 cm$^{-1}$ band to $^2T_{2g} \rightarrow {}^2T_{1g}({}^2A_{2g})$. It follows (dotted red line), that the crystal field splitting energy of low spin ferric iron ($10DqFe^{3+}_{LS}$) is ≈ 22200 cm$^{-1}$ at 40 GPa. The Racah parameter B = 600 cm$^{-1}$ is assumed consistent with reported values on HS $Fe^{3+}$ (Refs.[10, 44]).

Simultaneously with the anomalous behavior of the CT band, a new absorption peak centered at ~ 19200 cm$^{-1}$ appears in the spectrum (Fig. 2A). The new band blue-shifts with increasing pressure ($dv/dP$ = 46 cm$^{-1}$/GPa), which is characteristic of spin-allowed crystal field bands[10], but the pressure-dependent shift is much smaller as compared to that for LS $Fe^{2+}$ (88 cm$^{-1}$/GPa) in a similar crystal field[31]. The new band intensifies continuously at 40-46 GPa, indicating a growing population of the LS states. At P > 46 GPa, this band maintains an approximately constant intensity. We assign the band at ~ 19200 cm$^{-1}$ to the $^2T_{2g} \rightarrow {}^2T_{1g}({}^2A_{2g})$ transition in LS $Fe^{3+}$ as it is the lowest energy spin-allowed excited state (Fig. 4). We have a rough check on this assignment by accepting the Racah parameter B = 600 cm$^{-1}$, based on the reported values of B = 540-650 cm$^{-1}$ for ferric iron in an octahedral coordination[10, 44], allowing to estimate the crystal field splitting energy of LS $Fe^{3+}$ ($10DqFe^{3+}_{LS}$ ≈ 22200 cm$^{-1}$) as outlined in Figure 4 (dotted red line). At ambient conditions, crystal field splitting energies of HS $Fe^{3+}$ in iron oxides and hydroxides are in the range of 14000-16000 cm$^{-1}$ (Ref.[44]). Accepting a pressure shift of ~ 100 cm$^{-1}$/GPa (Ref.[45]), we obtain $10DqFe^{3+}_{HS}$ ≈ 18000-20000 cm$^{-1}$ at 40 GPa. Across the spin transition, *10Dq* increases abruptly as a result of a discontinuous Fe-O bond shortening.

For example, an increase in 10Dq of ~ 4000 cm$^{-1}$ is observed in the octahedrally-coordinated Fe$^{2+}$ in siderite (FeCO$_3$) across the spin transition[31]. Taking this into consideration we obtain *10DqFe$^{3+}_{LS}$* ≈ 22000-24000 cm$^{-1}$, close to what we have evaluated from the Tanabe-Sugano diagram (Fig. 4). This provides further support to the assignment of the new band to the $^2T_{2g} \rightarrow {}^2T_{1g}(^2A_{2g})$ transition in LS Fe$^{3+}$.

A decompression run shows that all the observed variations in optical absorbance are reversible (Fig. 2B), consistent with the spin transition scenario. However, a small hysteresis was observed for the low-to-high spin transition such that the $^2T_{2g} \rightarrow {}^2T_{1g}(^2A_{2g})$ band was present in the spectra down to ~ 37 GPa (Figs. 2B and 3). Interestingly, the new band reverses the sign of its frequency-pressure shift upon decompression below 44 GPa (Fig. 3), indicating an increase in the crystal field splitting energy at the LS Fe$^{3+}$ site upon the LS to HS transformation. This very unusual behavior suggests that the growing number of HS Fe$^{3+}$O$_6$-units induces local strain in the lattice, imposing additional stress to the yet untransformed LS Fe$^{3+}$ sites.

**Theoretical results**

The assignment of the 19200 cm$^{-1}$ band to the $^2T_{2g} \rightarrow {}^2T_{1g}(^2A_{2g})$ transition can be verified by computing the crystal field splitting energy of the LS *C*-site Fe$^{3+}$ using first-principles calculations. For the NAL phase with *C*-Fe$^{3+}$, standard DFT methods (e.g. PBE-GGA) can correctly predict an insulating LS state (see further discussions below) without including the Hubbard *U* correction. Consequently, PBE-GGA would give the correct orbital occupancy and electron charge density of the LS C-Fe$^{3+}$, as discussed in a previous study[46]. It should be emphasized, however, that the Hubbard *U* correction is still necessary to treat iron in other spin states and thus the spin transition in the NAL phase, similar to Bdgm. Using PBE-GGA, we optimize a 63-atom supercell with one single LS Fe$^{3+}$ substituting Al in the *C* site, namely, NaMg$_2$(Al$_{4.67}$Fe$_{0.33}$Si)O$_{12}$, to the cell volume *V* = 161.96 A$^3$/f.u., same as observed at *P* ≈ 40 GPa that marks the beginning of the HS-LS transition[35]. The resultant electronic structure is shown in Fig. 5, where the projected density of states (PDOS) onto each atomic species and the total DOS are plotted. The Fe *3d* orbitals form several narrow bands around the Fermi level $E_F$ (set to zero), giving rise to the Fe PDOS peaks indicated by the green (dotted) line. In the spin-up channel, the two Fe PDOS peaks below $E_F$ and the one Fe PDOS peak above $E_F$ correspond to the filled $t_{2g}$ and empty $e_g$ bands, respectively. The crystal field splitting, derived from the energy difference between the centers of the $t_{2g}$ and $e_g$ bands, is 2.70 eV (21800 cm$^{-1}$), in excellent agreement with 22000 cm$^{-1}$ derived from the Tanabe-Sugano diagram (Fig. 4) within the experiment uncertainty (200 cm$^{-1}$). Furthermore, DFT calculation shows that the crystal field splitting increases with pressure (not presented here), consistent with the blue shift of the $^2T_{2g} \rightarrow {}^2T_{1g}(^2A_{2g})$ band observed at $P > 40$ GPa shown in Fig. 3.

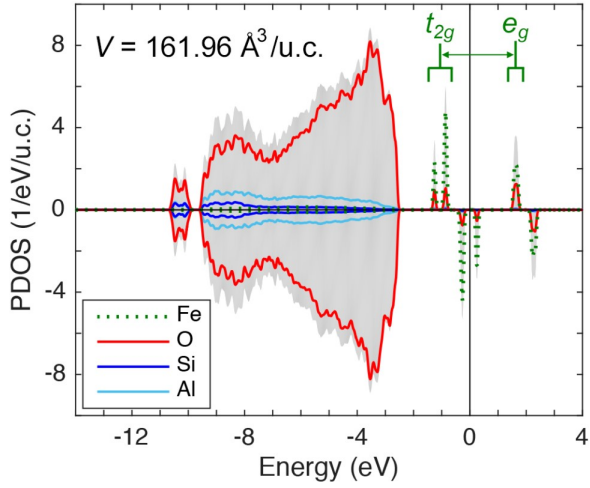

**Figure 5.** Projected (lines) and total (shade) density of states of $NaMg_2(Al_{4.67},Fe_{0.33},Si)O_{12}$ with LS $Fe^{3+}$ substituting Al in the $C$ site. At the unit cell $V = 161.96$ Å$^3$ ($P \approx 40$ GPa), the crystal field splitting (indicated by the double-headed arrow) is 2.70 eV (21800 cm$^{-1}$), in excellent agreement with 22000 cm$^{-1}$ as derived from the Tanabe-Sugano diagram in Figure 4.

Similar to the NAL phase, we determined the crystal field spitting of LS $Fe^{3+}$ at the $B$ site in $(Mg_{0.875}Fe_{0.125})(Si_{0.875}Fe_{0.125})O_3$ Bdgm with HS $Fe^{3+}$ at the dodecahedral $A$ site and LS $Fe^{3+}$ at the $B$ site. At $V = 145.99$ Å$^3$/cell, which marks the beginning of HS-LS transition observed in experiment at ~ 45 GPa (Ref.[15]), the separation between the $t_{2g}$ and $e_g$ bands is ~2.8 eV (22600 cm$^{-1}$) (Fig. 6), close to the crystal field splitting energy in LS NAL at 40 GPa (Fig. 4 and Fig. 5).

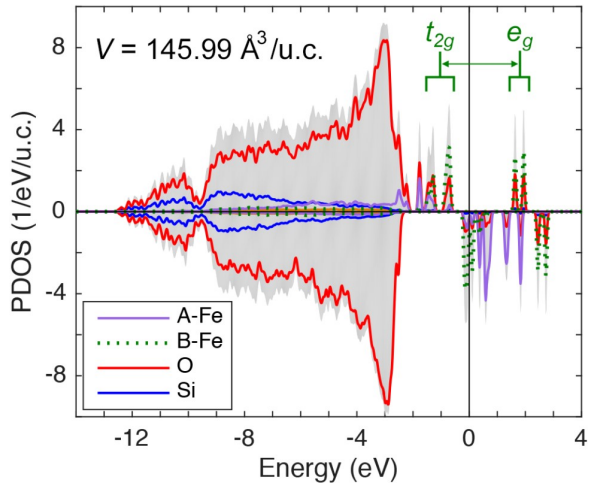

**Figure 6.** Projected (lines) and total (shade) density of states of $(Mg_{0.875}Fe_{0.125})(Si_{0.875}Fe_{0.125})O_3$ Bdgm. At the unit cell $V = 145.99$ Å$^3$ ($P \approx 40$ GPa), the crystal field splitting (indicated by the double-headed arrow) is 2.8 eV (22600 cm$^{-1}$).

**Discussion**

Spin transitions occur when the crystal field splitting energy of a cation ($10Dq$) exceeds its site-specific $d$-$d$ spin-pairing energy ($E_p$). This study allowed constraining the crystal field splitting energy of LS $Fe^{3+}$ at the octahedral site of NAL to $10DqFe^{3+}_{LS} = 22200$ cm$^{-1}$ at 40 GPa. This is in excellent agreement with the optical measurements of $E_p = 20600$-$22900$ cm$^{-1}$ of $Fe^{3+}$ in the strong octahedral field of $Fe(S_2CNR_2)_3$ at 1 atm[47]. Indeed, due to the nephelauxetic effect the spin-pairing energy of a cation in a crystal field is 70-85 % of the field-free cation value (29875 cm$^{-1}$ for field-free $Fe^{3+}$)[10]. Figure 7 compares the optically-derived $10Dq$ values of iron-bearing NAL as well as of other phases to the corresponding $E_p$ estimates for $Fe^{2+}$ and $Fe^{3+}$. Generally, crystal field splitting energies of HS and LS iron are consistent with the corresponding spin-pairing energy. For example, direct optical measurements of the crystal field splitting energy of HS and LS siderite show a straightforward behavior in that $10DqFe^{2+}_{HS} < E_p$ and $10DqFe^{2+}_{LS} > E_p$. Likewise, the available spectroscopic data on the electronic structure of octahedrally-coordinated $Fe^{3+}$, suggests that its spin-pairing energy is $E_p \approx 20000$-$23000$ cm$^{-1}$ (Fig. 7). Our DFT computations of the crystal field splitting energy in Bdgm (Fig. 6) yielded $10Dq_{Bdgm} = 22600$ cm$^{-1}$ for LS $Fe^{3+}$ at the octahedral site at $P \approx 40$ GPa. This value suggests that the theory-based view of the $B$-site $Fe^{3+}$ spin transition in Bdgm at $P > 40$ GPa (Refs.[24-26]) is consistent with the direct spectroscopic constraints on the spin-pairing energy in $Fe^{3+}O_6$-compounds (Fig. 7), such as determined in this work.

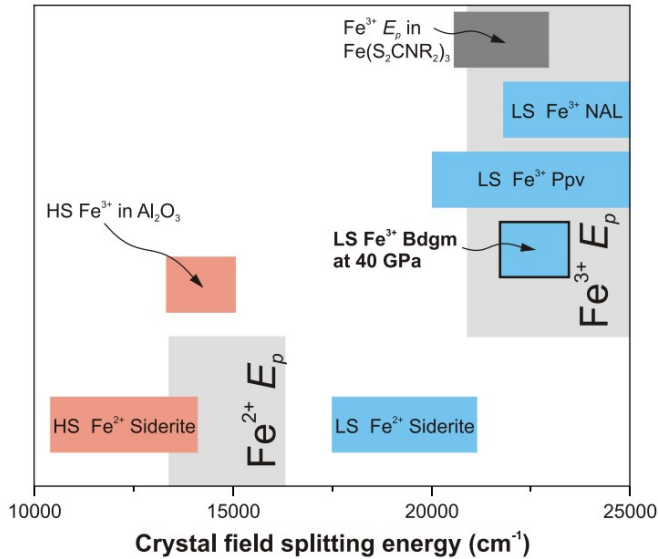

**Figure 7.** Crystal field splitting energies of iron in an octahedral crystal field of NAL (this study), post-perovskite (Ppv)[48], and siderite[31] as deduced from optical absorption measurements in DACs at 300 K. Red and blue horizontal bars show high and low spin values, respectively. The crystal field splitting energy of LS $Fe^{3+}$ in Bdgm is computed in this work (Fig. 6). Light grey vertical bars are approximate spin-pairing energies ($E_p$) of ferric and ferrous iron in a crystal field (estimated as 75-80 % of the spin-pairing energy for the field-free cations[10]. Dark grey bar is $E_p$ in Fe(S$_2$CNR$_2$)$_3$, which contains LS $Fe^{3+}$ in an octahedral field, as deduced from optical measurements across the temperature-induced spin transition[47]. The crystal field splitting energy of $Fe^{3+}$-bearing corundum is at 1 atm[10].

This empirical inference can be tested by direct optical observations of a spin transition in Bdgm, for which this work provides an important spectroscopic reference: the $^2T_{2g} \rightarrow {}^2T_{1g}({}^2A_{2g})$ band at ~ 17000-22000 cm$^{-1}$ is characteristic of octahedrally-coordinated LS $Fe^{3+}$. This is justified by the $d^5$ Tanabe-Sugano diagram as well as the similarity in *10Dq* of LS NAL and Bdgm. In further support, an absorption band centered at ~ 17500 cm$^{-1}$ is present in the absorption spectra of LS $Fe^{3+}$-bearing Mg$_{0.6}$Fe$_{0.4}$SiO$_3$ post-perovskite at 130 GPa and 300 K (Ref.[48]) which resembles NAL and Bdgm in the distribution of iron in the crystal structure[26, 27, 35]. It is clear, however, that the intensity of the $^2T_{2g} \rightarrow {}^2T_{1g}({}^2A_{2g})$ band is proportional to iron concentration at the octahedral site; thus, a substantial amount of LS $Fe^{3+}$ (~10 mol. %) is needed to detect the band. The overall solubility of $Fe^{3+}$ in Bdgm is enhanced by $Al^{3+}$ (*e.g.* Refs.[21, 49, 50]), but the predominant incorporation of $Al^{3+}$ on the octahedral *B* site limits heavily the concentration of $Fe^{3+}$ at this position. As a result, $Fe^{3+}$ in Bdgm with a mole fraction of $Al^{3+}$ > $Fe^{3+}$ is expected to occupy the *A* site mostly[18, 23]. A recent combined XES and Mőssbauer study of the iron spin state in Al-bearing Bdgm with almost all $Fe^{2+}$ and $Fe^{3+}$ in the *A* site has revealed that these iron ions remain in the HS state at lower-mantle pressures[30]. Therefore, all iron in Bdgm with $Al^{3+}$ > $Fe^{3+}$ is likely in the HS configuration at the conditions of Earth's lower mantle. It follows that the theoretically proposed spin transition scenario for $Fe^{3+}$ at the *B* site[13, 20, 24-28] may be achieved in samples with $Al^{3+}$ < $Fe^{3+}$, but available high-pressure optical absorption data on Bdgm with variable $Fe^{3+}/\Sigma Fe$ and $Al^{3+}$ content are very limited and inconclusive in addressing the spin state of iron[33, 34]. For instance, the $^2T_{2g} \rightarrow {}^2T_{1g}({}^2A_{2g})$ band was not detected in the absorption spectra of Mg$_{0.9}$Fe$_{0.1}$SiO$_3$ (11 ± 3 % $Fe^{3+}$)[33] and (Mg$_{0.892}$Fe$^{2+}_{0.059}$Fe$^{3+}_{0.042}$)(Si$_{0.972}$Al$_{0.028}$)O$_3$ Bdgm[34] up to the highest pressure of ~ 130 GPa. The presence/absence of $Fe^{3+}$ at the *B* site, however, has not been reported for these samples. Accordingly, it is unclear if the absence of the $^2T_{2g} \rightarrow {}^2T_{1g}({}^2A_{2g})$ band in these past studies is due to the exclusive HS iron configuration in the studied pressure range or due to the lack of $Fe^{3+}$ at the octahedral site. Systematic optical studies of Bdgm with variable compositions ($Fe^{2+}$, $Fe^{3+}$, $Al^{3+}$) at high pressure will help resolving this controversy in the Bdgm spin state.

**Acknowledgments**

This work was supported by the NSF Major Research Instrumentation program, NSF EAR-1015239, NSF EAR-1520648 and NSF EAR/IF-1128867, the Army Research Office (56122-CH-H), the Carnegie Institution of Washington and Deep Carbon Observatory. S.S.L. was partly supported by the Ministry of Education and Science of Russian Federation (No. 14B.25.31.0032) and by state assignment project No. 0330-2014-0013. H.H. is supported by the Ministry of Science and Technology of Taiwan under grant No. MOST 104-2112-M-008-005-MY3. J.F.L. acknowledges support from NSF Geophysics Program and CSEDI (EAR1446946, EAR1502594). A. F. G. was partly supported by the Chinese Academy of Sciences visiting professorship for senior international scientists (Grant No. 2011T2J20), Recruitment Program of Foreign Expert, the National Natural Science Foundation of China (grant number 21473211), and the Chinese Academy of Sciences (grant number YZ201524). Portions of this work were performed at GeoSoilEnviroCARS (Sector 13), Advanced Photon Source (APS), Argonne National Laboratory. GeoSoilEnviroCARS is supported by the National Science Foundation - Earth Sciences (EAR-1128799) and Department of Energy- GeoSciences (DE-FG02-94ER14466). This research used resources of the Advanced Photon Source, a U.S. Department of Energy (DOE) Office of Science User Facility operated for the DOE Office of Science by Argonne National Laboratory under Contract No. DE-AC02-06CH11357. We thank S. Fu and J. Yang for helping with polishing the crystals as well as Nicholas Holtgrewe for his comments on the earlier versions of this manuscript.

# References

1. Badro, J., *et al.* Iron partitioning in Earth's mantle: Toward a deep lower mantle discontinuity. *Science,* 300, 789-791 (2003).
2. Lin, J. F., Speziale, S., Mao, Z., Marquardt, H. Effects of the electronic spin transitions of iron in lower mantle minerals: iImplications for deep mantle geophysics and geochemistry. *Rev. Geophys.,* 51, 244-275 (2013).
3. Badro, J. Spin transitions in mantle minerals. *Ann. Rev. Earth Planet. Sci.,* 42, 231-248 (2014).
4. Goncharov, A. F., Struzhkin, V. V., Jacobsen, S. D. Reduced radiative conductivity of low-spin (Mg,Fe)O in the lower mantle. *Science,* 312, 1205-1208 (2006).
5. Keppler, H., Kantor, I., Dubrovinsky, L. S. Optical absorption spectra of ferropericlase to 84 GPa. *Am. Mineral.,* 92, 433-436 (2007).
6. Marquardt, H., *et al.* Elastic shear anisotropy of ferropericlase in Earth's lower mantle. *Science,* 324, 224-226 (2009).
7. Yang, J., Tong, X. Y., Lin, J. F., Okuchi, T., Tomioka, N. Elasticity of ferropericlase across the spin crossover in the Earth's lower mantle. *Sci. Rep.,* 5, 17188 (2015).
8. Vilella, K., Shim, S. H., Farnetani, C. G., Badro, J. Spin state transition and partitioning of iron: Effects on mantle dynamics. *Earth Planet. Sci. Lett.,* 417, 57-66 (2015).
9. Huang, C., Leng, W., Wu, Z. Q. Iron-spin transition controls structure and stability of LLSVPs in the lower mantle. *Earth Planet Sci. Lett.,* 423, 173-181 (2015).
10. Burns, R. G. *Mineralogical applications of crystal field theory*, (Cambridge University Press, U.K., 1993).
11. Badro, J., *et al.* Electronic transitions in perovskite: Possible nonconvecting layers in the lower mantle. *Science,* 305, 383-386 (2004).
12. Li, J., *et al.* Electronic spin state of iron in lower mantle perovskite. *Proc. Natl. Acad. Sci. U.S.A.,* 101, 14027-14030 (2004).
13. Zhang, F. W. & Oganov, A. R. Valence state and spin transitions of iron in Earth's mantle silicates. *Earth Planet. Sci. Lett.,* 249, 436-443 (2006).
14. McCammon, C., *et al.* Low-spin $Fe^{2+}$ in silicate perovskite and a possible layer at the base of the lower mantle. *Phys. Earth Planet. Inter.,* 180, 215-221 (2010).
15. Catalli, K., *et al.* Spin state of ferric iron in $MgSiO_3$ perovskite and its effect on elastic properties. *Earth Planet. Sci. Lett.,* 289, 68-75 (2010).
16. Catalli, K., *et al.* Effects of the $Fe^{3+}$ spin transition on the properties of aluminous perovskite-New insights for lower-mantle seismic heterogeneities. *Earth Planet. Sci. Lett.,* 310, 293-302 (2011).
17. Lin, J. F., *et al.* Electronic spin states of ferric and ferrous iron in the lower-mantle silicate perovskite. *Am. Mineral.,* 97, 592-597 (2012).
18. Kupenko, I., *et al.* Oxidation state of the lower mantle: In situ observations of the iron electronic configuration in bridgmanite at extreme conditions. *Earth Planet. Sci. Lett.,* 423, 78-86 (2015).
19. Dorfman, S. M., *et al.* Composition dependence of spin transition in (Mg,Fe)$SiO_3$ bridgmanite. *Am. Mineral.,* 100, 2246-2253 (2015).
20. Mohn, C. E. & Trønnes, R. G. Iron spin state and site distribution in $FeAlO_3$-bearing bridgmanite. *Earth Planet. Sci. Lett.,* 440, 178-186 (2016).
21. McCammon, C. Perovskite as a possible sink for ferric iron in the lower mantle. *Nature,* 387, 694-696 (1997).
22. Hummer, D. R. & Fei, Y. W. Synthesis and crystal chemistry of $Fe^{3+}$-bearing (Mg,$Fe^{3+}$)(Si,$Fe^{3+}$)$O_3$ perovskite. *Am. Mineral.,* 97, 1915-1921 (2012).
23. Potapkin, V., *et al.* Effect of iron oxidation state on the electrical conductivity of the Earth's lower mantle. *Nat. Commun.,* 4, 1427 (2013).
24. Hsu, H., Umemoto, K., Blaha, P., Wentzcovitch, R. M. Spin states and hyperfine interactions of iron in (Mg,Fe)$SiO_3$ perovskite under pressure. *Earth Planet. Sci. Lett.,* 294, 19-26 (2010).
25. Hsu, H., Blaha, P., Cococcioni, M., Wentzcovitch, R. M. Spin-state crossover and hyperfine interactions of ferric iron in $MgSiO_3$ perovskite. *Phys. Rev. Lett.,* 106, 118501 (2011).
26. Yu, Y. G. G., Hsu, H., Cococcioni, M., Wentzcovitch, R. M. Spin states and hyperfine interactions of iron incorporated in $MgSiO_3$ post-perovskite. *Earth Planet. Sci. Lett.,* 331, 1-7 (2012).
27. Hsu, H., Yu, Y. G. G., Wentzcovitch, R. M. Spin crossover of iron in aluminous $MgSiO_3$ perovskite and post-perovskite. *Earth Planet. Sci. Lett.,* 359, 34-39 (2012).

28. Hsu, H. & Wentzcovitch, R. M. First-principles study of intermediate-spin ferrous iron in the Earth's lower mantle. *Phys. Rev. B,* 90, 195205 (2014).
29. Dyar, M. D., Agresti, D. G., Schaefer, M. W., Grant, C. A., Sklute, E. C. Mőssbauer spectroscopy of Earth and planetary materials. *Annu. Rev. Earth Planet. Sci.,* 34, 83-125 (2006).
30. Lin, J.-F., *et al.* High-spin $Fe^{2+}$ and $Fe^{3+}$ in single-crystal aluminous bridgmanite in the lower mantle. *Geophys. Res. Lett.,* 6952-6959 (2016).
31. Lobanov, S. S., Goncharov, A. F., Litasov, K. D. Optical properties of siderite ($FeCO_3$) across the spin transition: Crossover to iron-rich carbonates in the lower mantle. *Am. Mineral.,* 100, 1059-1064 (2015).
32. Lobanov, S. S., Holtgrewe, N., Goncharov, A. F. Reduced radiative conductivity of low spin $FeO_6$-octahedra in $FeCO_3$ at high pressure and temperature. *Earth Planet. Sci. Lett.,* 449, 20-25 (2016).
33. Goncharov, A. F., Haugen, B. D., Struzhkin, V. V., Beck, P., Jacobsen, S. D. Radiative conductivity in the Earth's lower mantle. *Nature,* 456, 231-234 (2008).
34. Keppler, H., Dubrovinsky, L. S., Narygina, O., Kantor, I. Optical absorption and radiative thermal conductivity of silicate perovskite to 125 Gigapascals. *Science,* 322, 1529-1532 (2008).
35. Wu, Y., *et al.* Spin transition of ferric iron in the NAL phase: Implications for the seismic heterogeneities of subducted slabs in the lower mantle. *Earth Planet. Sci. Lett.,* 434, 91-100 (2016).
36. Ricolleau, A., *et al.* Phase relations and equation of state of a natural MORB: Implications for the density profile of subducted oceanic crust in the Earth's lower mantle. *Journal of Geophysical Research,* 115, B08202 (2010).
37. Walter, M. J., *et al.* Deep mantle cycling of oceanic crust: Evidence from diamonds and their mineral inclusions. *Science,* 334, 54-57 (2011).
38. Goncharov, A. F., Beck, P., Struzhkin, V. V., Haugen, B. D., Jacobsen, S. D. Thermal conductivity of lower-mantle minerals. *Phys. Earth Planet. Inter.,* 174, 24-32 (2009).
39. Giannozzi, P., *et al.* QUANTUM ESPRESSO: a modular and open-source software project for quantum simulations of materials. *J. Phys. Cond. Matter,* 21, 395502 (2009).
40. Perdew, J. P., Burke, K., Ernzerhof, M. Generalized gradient approximation made simple. *Phys. Rev. Lett.,* 77, 3865-3868 (1996).
41. Vanderbilt, D. Soft self-consistent pseudopotentials in a generalized eigenvalue formalism. *Phys. Rev. B,* 41, 7892-7895 (1990).
42. Garrity, K. F., Bennett, J. W., Rabe, K. M., Vanderbilt, D. Pseudopotentials for high-throughput DFT calculations. *Comp. Mater. Sci.,* 81, 446-452 (2014).
43. Mattson, S. M. & Rossman, G. R. Identifying characteristics of charge transfer transitions in minerals. *Phys. Chem. Miner.,* 14, 94-99 (1987).
44. Sherman, D. M. & Waite, T. D. Electronic spectra of $Fe^{3+}$ oxides and oxide hydroxides in the near IR to near UV. *Am. Mineral.,* 70, 1262-1269 (1985).
45. Mao, H. K. & Bell, P. M. Crystal-field effects of ferric iron in goethite and lepidocrocite: band assignments and geochemical application at high pressure. *Carnegie Inst. Washington Yearb.,* 1973, 502-507 (1974).
46. Hsu, H., Umemoto, K., Cococcioni, M., Wentzcovitch, R. M. The Hubbard U correction for iron-bearing minerals: A discussion based on $(Mg,Fe)SiO_3$ perovskite. *Phys. Earth Planet. Inter.,* 185, 13-19 (2011).
47. Ewald, A. H., Martin, R. L., Sinn, E., White, A. H. Electronic equilibrium between the $^6A_1$ and $^2T_2$ states in iron(III) dithio chelates. *Inorg. Chem.,* 8, 1837-1846 (1969).
48. Lobanov, S. S., Holtgrewe, N., Lin, J. F., Goncharov, A. F. Radiative conductivity and abundance of post-perovskite in the lowermost mantle. *arXiv,* arXiv:1609.06996 (2016).
49. Sinmyo, R., Hirose, K., Muto, S., Ohishi, Y., Yasuhara, A. The valence state and partitioning of iron in the Earth's lowermost mantle. *Journal of Geophysical Research,* 116, B07205 (2011).
50. Lauterbach, S., McCammon, C. A., van Aken, P., Langenhorst, F., Seifert, F. Mossbauer and ELNES spectroscopy of $(Mg,Fe)(Si,Al)O_3$ perovskite: a highly oxidised component of the lower mantle. *Contrib. Mineral. Petrol.,* 138, 17-26 (2000).